\documentclass[aps,prd, nofootinbib,
preprintnumbers,showkeys, preprint,
superscriptaddress]{revtex4-1}

\usepackage{graphicx}
\usepackage{tikz}
\usetikzlibrary{shapes.misc}

\tikzset{
    cross/.pic = {
    \draw[rotate = 45, very thick] (-#1,0) -- (#1,0);
    \draw[rotate = 45, very thick] (0,-#1) -- (0, #1);
    }
}
\usepackage{physics}
\usetikzlibrary{decorations.markings}
\usepackage{subcaption}
\usepackage{dcolumn}
\usepackage{bm}
\usepackage{amsmath}
\usepackage{wasysym}
\usepackage{float}
\usepackage{multirow}
\usepackage{xcolor}
\usepackage{scrextend}
\usepackage[colorlinks=true, 
pdfstartview=FitV, 
bookmarks=true, 
bookmarksnumbered=true, 
breaklinks]{hyperref}
\usepackage{color}
\usepackage[normalem]{ulem} %strikethrough
\definecolor{blue}{rgb}{0.0, 0.0, 1.0}
\definecolor{red}{rgb}{1.0, 0.0, 0.0}
\definecolor{royalblue}{rgb}{0.0, 0.14, 0.4}
\hypersetup{linkcolor=royalblue, 
	citecolor=blue, 
	urlcolor=royalblue}

\def\orcid#1{\kern .08em\href{https://orcid.org/#1}{\includegraphics[keepaspectratio,width=0.7em]{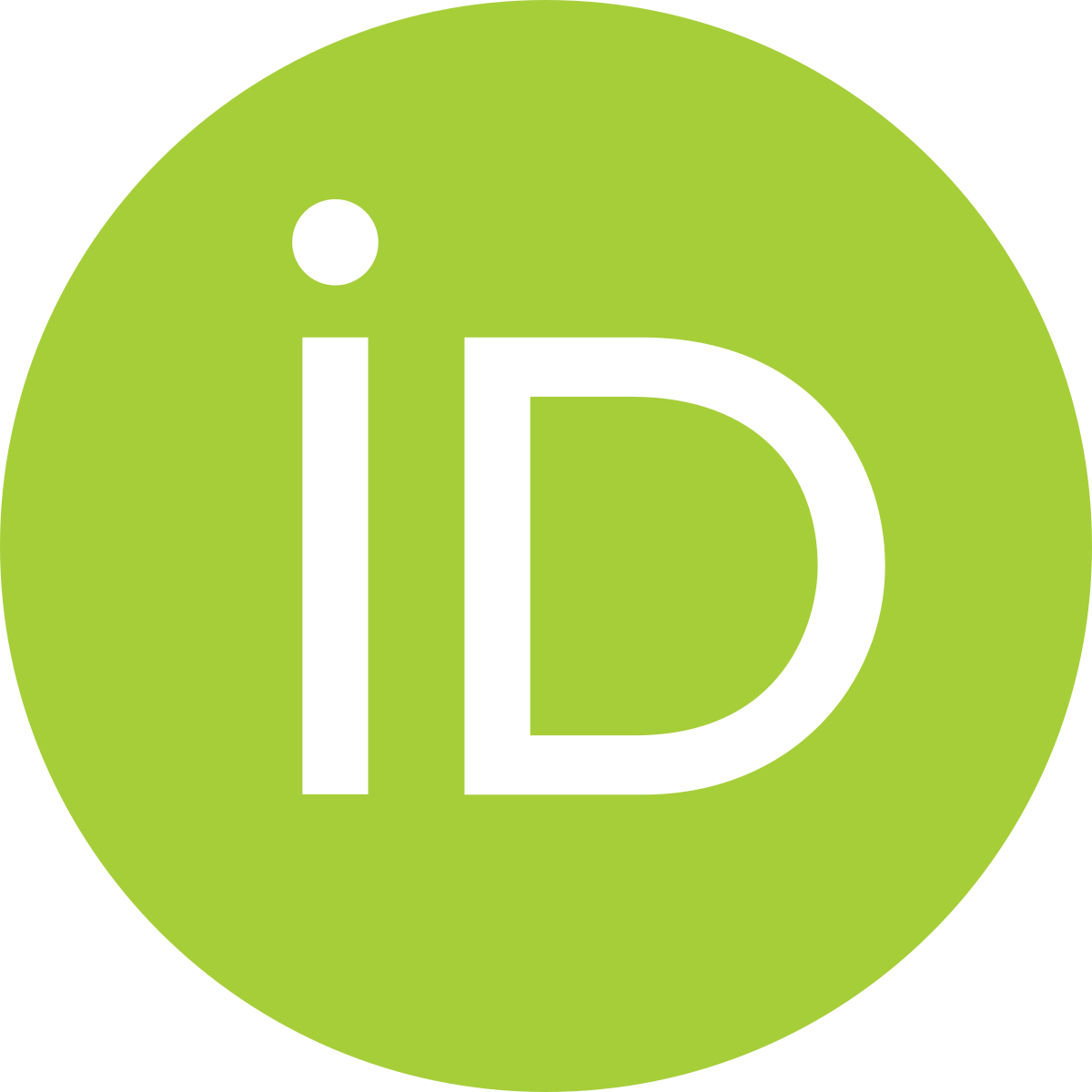}}}

\definecolor{ttcolor}{RGB}{240,248,255}
\definecolor{btcolor}{RGB}{144,238,144}
\definecolor{tbcolor}{RGB}{216,191,216}
\definecolor{bbcolor}{RGB}{255,228,225}

\begin{document}
\title{Interpretation of near-threshold peaks using the method of independent S-matrix poles}
%\title{What we miss out using the effective range expansion in the $P_\psi^N(4312)^+$ signal analysis}
% \title{Differentiating identical line shapes of the scattering amplitude}
%\title{Ambiguity in the interpretation of near-threshold signals}
%\title{On the interpretation of near-threshold line shapes}

\author{Leonarc Michelle Santos\orcid{0000-0003-0444-5544}}
\email[]{lsantos@up.edu.ph}
\affiliation{National Institute of Physics, University of the Philippines Diliman, Quezon City 1101, Philippines}

\author{Denny Lane B. Sombillo\orcid{0000-0001-9357-7236}}
%\email[]{sombillo@rcnp.osaka-u.ac.jp}
\email[]{dbsombillo@up.edu.ph}
\affiliation{National Institute of Physics, University of the Philippines Diliman, Quezon City 1101, Philippines}
%\affiliation{Research Center for Nuclear Physics (RCNP), Osaka University, Ibaraki, Osaka 567-0047, Japan}

\date{\today}
\begin{abstract}
    \noindent
    We propose a model-independent analysis of near-threshold enhancements using independent S-matrix poles. In this formulation, we constructed a Jost function with controllable zeros to ensure that no poles are generated on the physical Riemann sheet. We show that there is a possibility of misinterpreting the observed near-threshold signals if one utilized a limited parametrization and restrict the analysis to only one element of the S-matrix. Specifically, there is a possibility of the emergence of ambiguous pair of poles which are singularities of the full S-matrix but may not manifest in one of its elements. For a more concrete discussion, we focused on an effective two-channel scattering where the full S-matrix is a $2\times2$ matrix. We apply our method to the coupled two-channel analysis of the $P_\psi^N(4312)^+$ and found that the compact pentaquark interpretation cannot be ruled out yet.
\end{abstract}		

\maketitle
\section{Introduction}
% \textcolor{red}{Color confinement is one of the hardest and longest standing problems in hadron physics. The dynamical description of quarks and gluons becomes non trivial at low-energy due to the non-perturbative nature of strong interactions. At best, one can only rely on the brute force numerical calculations of lattice QCD or through the use of phenomenological models to gain some insights. 
% An alternative way to get a hint of how confinement work is through the bottom-top approach in spectroscopy \cite{Novel}. That is, we identify which of the observed signals qualify to be part of the hadron spectrum. However, classifying signals itself posit a new set of problems. That is, some of the observed enhancements in the scattering or invariant mass distributions might not be due to the excitation of the bound quark system in a hadron. }
One of the active areas of investigation in hadron spectroscopy is the interpretation of near-threshold phenomena \cite{Olsen2018,Guo2018,Oller:2019opk,TriangleAndCusp,Mai:2022eur,ALBALADEJO2022103981}. 
In 2019, an updated analysis of the $\Lambda^0_b \to J/\psi pK^-$ decays based on Runs 1 and 2 of the LHCb collaboration was presented in Ref.~\cite{virtualLHCbcollab}. They observed the narrow pentaquark state $P_\psi^N(4312)^+$ (then called the $P_c(4312)^+$) together with the two-peak structure of the $P_c(4450)^+$ resonance which was not present in their initial analysis in Ref.~\cite{LHCB2015}. These newly observed resonances have narrow decay widths and are below the $\Sigma_c\Bar{D}$ or $\Sigma_c\Bar{D}^*$, a typical signature of molecules. Ref.~\cite{virtualLHCbcollab} had concluded that $P_\psi^N(4312)^+$ is a virtual state with the $\Sigma^+_c\bar{D}^0$ threshold being within its extent. A similar parametrization study Ref.~\cite{Pc4312} and a deep learning approach Ref.~\cite{DeepLearningExHad} had the same conclusion for the $P_\psi^N(4312)^+$ signal.

Other interpretations are possible and were done in different studies. For example, in Ref.~\cite{QCDsumrules} the $P_\psi^N(4312)^+$ resonance is favored to have a molecular structure using the QCD sum rules formalism. It was found to be the $[\Sigma_c^{++}\Bar{D}^-]$ bound state with $J^P=1/2^-$. The molecular picture and same quantum number is favored as well in Ref.~\cite{chengliumolecular} where they studied the mass and decay properties of $P_\psi^N(4312)^+$ using isospin breaking effects and rearrangement decay properties. Ref.~\cite{KinematicsPhysRevLett.124.072001} used a coupled-channel formalism and ends up with the same conclusion.

In Ref.~\cite{matuschek2021nature}, it was argued that if range corrections can be neglected, virtual states are molecular in nature and hence the studies cited above poses no contradiction. As can be observed, the molecular picture is favored by most studies. However, we still cannot dismiss the compact pentaquark picture since information about quantum numbers and decay properties is lacking experimentally. In fact a model based approach in Ref.~\cite{ALI2019365compact}, the resonance was studied under the compact diquark model as a hidden-charm diquark-diquark-antiquark baryon with $J^P = 3/2^+$.

Until we settle the quantum numbers and decay properties of the $P_c$ resonances of the $\Lambda_b^0 \to J/\psi pK^-$, we cannot completely rule out the compact nature of the $P_\psi^N(4312)^+$. A good way to investigate this resonance with some of its properties still ambiguous is by using a minimally biased bottom-up approach study \cite{ALBALADEJO2022103981}. The $S$-matrix is a good tool to use in a bottom-up approach study as it can be constructed without any details of the interacting potential. We only need  to impose analyticity, and hermiticity and unitarity below the first threshold. Given these three mathematical restrictions, we can reproduce the scattering amplitude from experiment by identifying the optimal placement and number of poles in the scattering process. Finally, we can look for a theoretical model that can reproduce the same analytic properties of the constructed $S$-matrix.

%The $S$-matrix is a capable tool to use with very minimal model prescription. The nature of its poles and position in the Riemann sheets have correspondence with resonances (see Ref.~\cite{Badalyan1982} for a comprehensive review). We need only to impose analyticity, and hermiticity and unitarity below the first threshold.

In this paper, we show that some arrangements of poles may not be accommodated by the usual amplitude parametrizations such as the effective range expansion.
%In this paper, we show that certain pole location arrangements can be obscured when constructing scattering amplitudes under certain approximations.
Specifically, there are combinations of poles which will not manifest in the line shape of the elastic scattering amplitude.  
%Specifically, if we add these particular poles which we will be calling ambiguous pair poles, they may have no effective contribution to the elastic scattering amplitude. 
This, in turn, opens up the possibility of having a pole structure that caters the compact nature of $P_\psi^N(4312)^+$. The difficulty of capturing these subtle pole configurations may arise due to the contamination of coupled channel effects.
%To elaborate, the analysis of near-threshold phenomena is made more difficult due to the contamination of coupled channel effects. 
For example, a weakly interacting final state hadrons (higher mass channel) may have a virtual state pole that can be displaced away from the real energy axis due to coupling with the lower mass channel. If there is a resonance that is strongly coupled to the lower mass channel, then the displaced virtual state pole and the shadow pole of the resonance may have a cancellation effect in the elastic transition amplitude. One can invoke the pole-counting method to interpret a near-threshold pole with an accompanying shadow pole as non-molecular  \cite{Morgan1992,MorganPennington1991,morgan1993decay}.
 %can mislead the interpretation of a compact state because the pole structure can be more complicated than what it was initially thought of. 
 %As the main thesis of this work, we will be showing that the scattering amplitude of the $P_\psi^N(4312)^+$ signal can be reproduced by a pole structure of a compact state as interpreted by the pole counting method 
 In this work, we propose to use the independent S-matrix poles to accommodate all possible interpretations of the observed near-threshold enhancements.

The content of this paper is organized as follows: In section~\ref{sec:II}, we review the formalism of the $S$-matrix and show how one can construct an S-matrix using independent poles via the uniformization scheme introduced in Refs.~\cite{Newton, Kato1965, Yamada2020, Yamada2021, yamada2022near}. In section~\ref{sec:III}, we discuss how identical line shapes arises. We show that the ambiguity can be resolved by adding the contribution of the off diagonal $T_{21}$ channel. As an application, we investigate the $P_\psi^N(4312)^+$ signal and show that its $T_{11}$ line shape can take a $1$-pole configuration or $3$-pole configuration. %A more extensive analysis on this will be done in a different work. 
In section~\ref{sec:V}, we give our conclusion and outlook for future works.

\section{Formalism}\label{sec:II}
\hspace{\parindent}
The $S$-matrix is an operator that describes the interaction of a scattering process. In momentum space, one could decompose the $S$-matrix in terms of the non-interacting terms and interacting terms as \cite{Taylor}
\begin{align}\label{eqn:Staylor}
\bra{\mathbf{p'}}S\ket{\mathbf{p}} = \delta^3(\mathbf{p'}-\mathbf{p}) + \frac{i}{2\pi m}\delta(E_{p'}-E_p) f(\mathbf{p'}\leftarrow\mathbf{p}) 
\end{align}
where $f(\mathbf{p'}\leftarrow\mathbf{p})$ is the scattering amplitude from momentum $\mathbf{p}$ to $\mathbf{p'}$ and the factors of the interacting term may vary depending on the literature. The scattering amplitude $f(\mathbf{p'}\leftarrow\mathbf{p})$ is related to the cross section via
\begin{align}\label{eqn:crosssec}
\frac{\mathrm{d}\sigma}{\mathrm{d}\Omega}(\mathbf{p}\leftarrow\mathbf{p_0}) = |f(\mathbf{p}\leftarrow\mathbf{p_0})|^2.
\end{align}
In principle, we could use equations \eqref{eqn:Staylor} and \eqref{eqn:crosssec} to construct a parametrized $S$-matrix to fit it in the measured cross section data from experiments. The peaks from such data are characterized by the pole singularities of the $S$-matrix and it is indicative of the nature of the intermediate particles. 

In practice, the usual treatment of peaks in the scattering cross section is to utilize the Breit-Wigner parametrization to extract the mass and width of the resonance. This approach works very well if the peaks are far from any threshold and the widths are narrow. However, most of the recently observed peaks occur very close to some two-hadron threshold, where the peaks are no longer a reliable information to quote the mass of the observed state. Moreover, coupled-channel effects can no longer be ignored if the peaks are close to the thresholds. Unlike the complex energy plane of a single-channel system, one has to probe deeper into the multiple Riemann sheets of the energy complex plane of a multi-channel scattering. Specifically, the peaks observed may correspond to different pole arrangements in different Riemann sheets. 

In a one-channel scattering, we are interested in singularities of the complex momentum $p$ - plane. The poles on its positive imaginary axis correspond to bound states and the poles below its real axis may correspond to resonances. With the relativistic energy-momentum relation
$p^\mu p_\mu = E^2 - \Vec{p}\cdot\Vec{p} = m^2$,
the complex momentum $p$-plane transforms into two Riemann sheets of the complex energy $E$. The first Riemann sheet (top sheet) of $E$, which we call the physical sheet, corresponds to the upper half plane of $p$. The importance of physical sheet is that the scattering region lies on this complex energy plane. The scattering region corresponds to the energy axis used in plotting scattering observables. The second Riemann sheet (bottom sheet) of $E$, which we call the unphysical sheet, corresponds to the lower half plane of $p$. Due to causality, no other singularities should be present in the physical energy sheet aside from bound state poles and a branch point at the threshold \cite{Kampen1953, Kampen1953_1}.

Accordingly, in a two-channel scattering, we get four Riemann sheets (see \cite{Rakityansky2022} for an in-depth discussion). Only poles closest to the scattering region are relevant in the description of scattering data. In this paper, we  used the notation of Pearce and Gibson in \cite{PearceGibson} in labeling our Riemann sheets. We label the sheets as $[XY]$ where the string can be $t$ or $b$ to denote a top sheet or bottom sheet and the order of character denotes the channel. For example. the sheet $[bt]$ corresponds to the bottom sheet of the first channel and top sheet of the second channel. The correspondence of Pearce and Gibson's notation with the more commonly used notation of Frazer and Hendry \cite{Frazer1964} is listed in Table~\ref{tab:notation}.

\begin{table}[h!]
    \centering
    \begin{tabular}{|c|c|c|}
    \hline
         Frazer and Hendry & Pearce and Gibson & Topology in complex $E$ \\
    \hline
         I & $[tt]$ & $\theta_1 \in (0,2\pi)$; $\theta_2 \in (0,2\pi)$  \\
    \hline
        II & $[bt]$ & $\theta_1 \in (2\pi, 4\pi)$; $\theta_2 \in (0,2\pi)$ \\
    \hline
        III & $[bb]$ & $\theta_1 \in (2\pi,4\pi)$; $\theta_2 \in (2\pi,4\pi)$ \\
    \hline
        IV & $[tb]$ & $\theta_1 \in (0,2\pi)$; $\theta_2 \in (2\pi,4\pi)$ \\
    \hline
    \end{tabular}
    \caption{Riemann sheet notation. In this work, we will follow Pearce and Gibson's notation. The index correspond to the channel number and $t(b)$ denotes top(bottom) sheet.}
    \label{tab:notation}
\end{table}

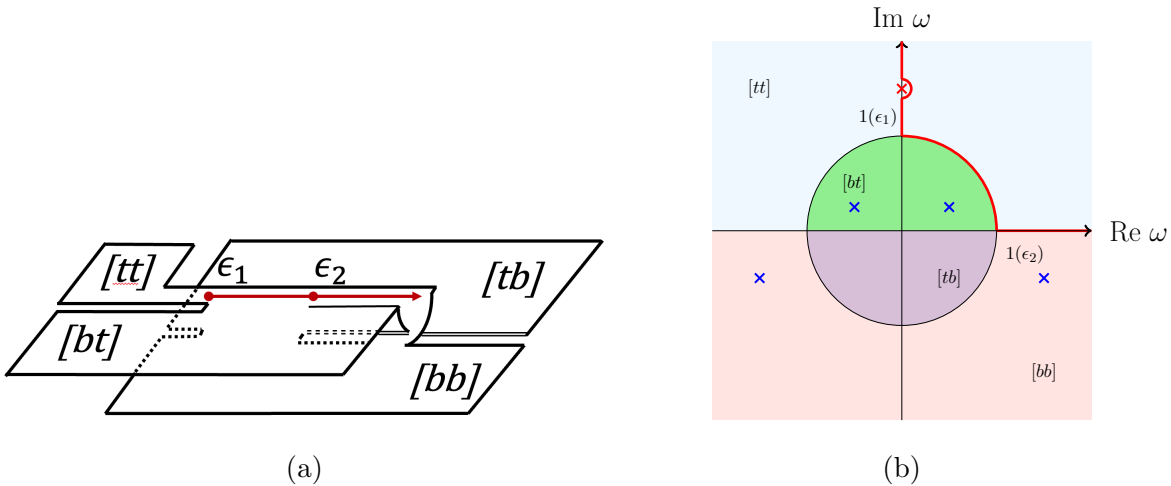
\begin{figure}[h!]
        \centering
        \begin{subfigure}[b]{0.5\columnwidth}
            \centering
            \includegraphics[scale=0.3]{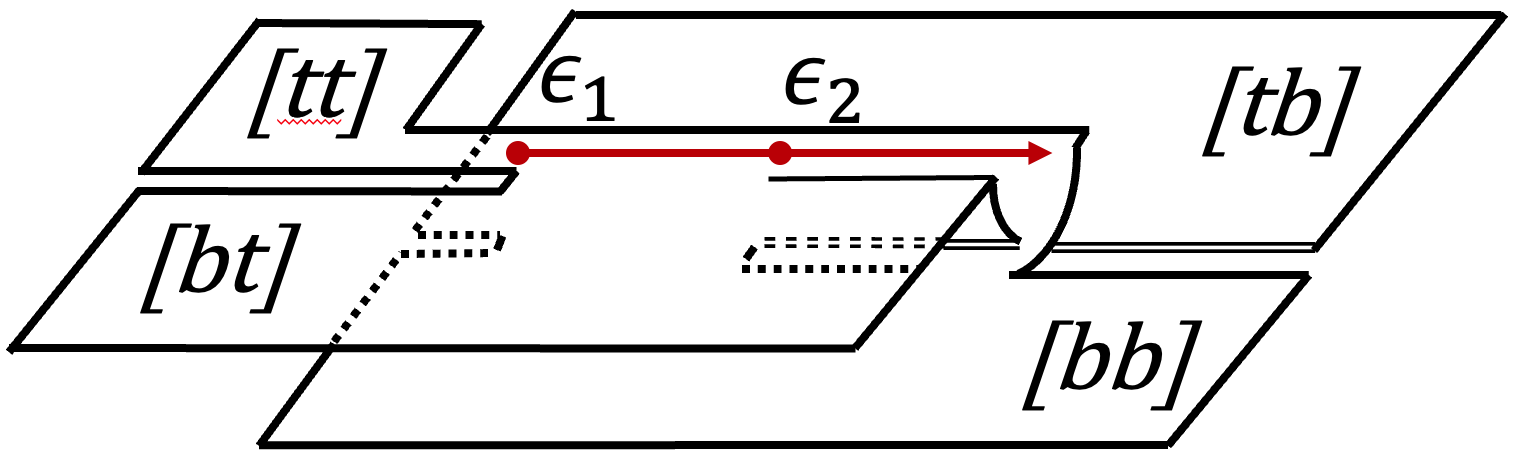} 
        \caption[]{}\label{1aEplane}
        
        \end{subfigure}
        %\hfill
        \begin{subfigure}[b]{0.45\columnwidth}  
            \resizebox{\textwidth}{!}{
                \begin{tikzpicture}
                    \node[font = {\Large}] at (4,8.5) {Im $\omega$};
                    \node[font = {\Large}] at (9,4) {Re $\omega$};
                    \node[font = {\Large}, opacity = 0] at (-1,4) {Re $\omega$};
                    \fill[bbcolor] (0,0) rectangle (8,4);
                    \node[fill = bbcolor, font = {\small}] at (7,1) {$[bb]$};
                    \fill[ttcolor] (0,4) rectangle (8,8);
                    \node[fill = ttcolor, font = {\small}] at (1,7) {$[tt]$};
                    \fill[tbcolor] (2,4) arc (180:360:2cm);
                    \node[fill = tbcolor, font = {\small}] at (5,3) {$[tb]$};
                    \fill[btcolor] (6,4) arc (0:180:2cm);
                    \node[fill = btcolor, font = {\small}] at (3,5) {$[bt]$};
                    
                    \draw (4,4) circle (2cm);
                    \draw[red!100, ultra thick] (6,4) arc (0:90:2cm);
                    \node[fill = ttcolor, font = {\small}] at (3.5, 6.4) {$1(\epsilon_1)$};
                    \node[fill = bbcolor, font = {\small}] at (6.6, 3.5) {$1(\epsilon_2)$}; 
                    \draw[red!100, ultra thick] (6,4) -- (8,4);
                    \draw[decoration={markings,mark=at position 1 with {\arrow[ultra thick]{>}}},
                        postaction={decorate}] (4,0) -- (4,8);
                    \draw[decoration={markings,mark=at position 1 with {\arrow[ultra thick]{>}}},
                        postaction={decorate}] (0,4) -- (8,4);
                    
                    \draw[red!100, ultra thick] (4,6) -- (4,6.8);
                    \path (4,7) pic[red] {cross=4pt};
                    \draw[red!100, ultra thick] (4,6.8) arc (270:450:0.20cm);
                    \draw[red!100, ultra thick] (4,7.2) -- (4,8);
                    
                    \path (3,4.5) pic[blue] {cross=4pt};
                    \path (5,4.5) pic[blue] {cross=4pt};
                    \path (1,3) pic[blue] {cross=4pt};
                    \path (7,3) pic[blue] {cross=4pt};
                \end{tikzpicture}
            }
                \caption[]{}\label{1bomegaplane}  
            
        \end{subfigure}
        \caption[]
        {The relevant regions of the four Riemann sheets in a two-channel scattering. (a) The energy complex plane. (b) The uniformized variable $\omega$ plane mapped from the complex energy $E$-plane using uniformization.}\label{fig:RS_2channels}
    \end{figure}

% \begin{figure}[h!]
% \includegraphics[width=0.5\columnwidth]{00_RS_2channels.jpg}
% \caption{The relevant regions of the four Riemann sheets in a two-channel scattering. (a) The energy complex plane. The scattering region represented by a ray with two dots is lying on the physical sheet $[tt]$. The portion of the $[bt]$ ($[bb]$) sheet above the second threshold $\epsilon_2$ is connected to the $[tb]$ ($[tt]$) sheet. The $[tt]$ is cut open to expose the relevant region of the $[tb]$ sheet. (b) The uniformized variable $\omega$ plane. We will show in section~\ref{sec:uniformization}}
% \label{fig:RS_2channels}
% \end{figure}

 % The portion of the $[bt]$ ($[bb]$) sheet above the second threshold $\epsilon_2$ is connected to the $[tb]$ ($[tt]$) sheet. The $[tt]$ is cut open to expose the relevant region of the $[tb]$ sheet. 

We show in Figure~\ref{1aEplane} an illustration of the four Riemann sheets in a two-channel scattering. The scattering region is represented by the red ray with two dots (branch points) lying on the physical sheet $[tt]$. It is directly connected with the lower halves of the $[bt]$ and $[bb]$ sheets. Crossing the branch cut between the first energy threshold $\epsilon_1$ and second energy threshold $\epsilon_2$, will send us to the $[bt]$ sheet. On the other hand, if we cross the branch cut above $\epsilon_2$, we end up in the $[bb]$ sheet. Energy poles found on these two sheets are quoted with negative imaginary parts. Consequently, these poles fit the description of unstable quantum states since their negative imaginary parts can reproduce the expected exponential decay of unstable states. On the other hand, energy poles on the $[tb]$ sheet are quoted with positive imaginary parts since only the upper half plane of the $[tb]$ sheet affect the scattering region. This gives rise to an exponential increase in time which does not correspond to any quantum state.

% Given an initial state $\ket{\phi}$ and a final state $\ket{\chi}$, the quantity of interest is the probability that the initial state $\ket{\phi}$ undergoing a scattering will emerge as the final state $\ket{\chi}$, i.e., $w(\chi \leftarrow \phi)$. This probability is computed using the equation $w(\chi \leftarrow \phi) = |\bra{\chi}S\ket{\phi}|^2$. We see that the probability amplitude for the process $\chi \leftarrow \phi$ is an element of the $S$-matrix describing the general scattering process.
\subsection{Analytic Structure of a two-channel S-matrix}
Without loss of generality, we can focus on an effective two-channel scattering.  The full two-channel transition is described by a $2\times2$ $S$-matrix whose elements are given by
\begin{align}\label{eq:Sdiag}
S_{11}(p_1,p_2)&=
        \dfrac{D(-p_1,p_2)}
        {D(p_1, p_2)}; \quad S_{22}(p_1,p_2)=
        \dfrac{D(p_1,-p_2)}
        {D(p_1, p_2)} 
\end{align}
and
\begin{align}\label{eq:Soffdiag}
    S_{12}^2=S_{11}S_{22}-
        \text{det}S    
\end{align}
where $D(p_1,p_2)$ is the Jost function \cite{LeCouter1960, Newton1961, Newton1962RelaxCons}. The subscripts correspond to the channel index with $1$ representing the lower mass channel, and $2$ for the higher mass. Causality requires that the $S$-matrix be analytic up to the branch points and poles \cite{Kampen1953_1, Kampen1953}. These singularities are related to the details of scattering. Branch cuts are dictated by kinematics while the poles are dynamical in nature. 
Depending on the location of the pole on the Riemann sheet, it may in general correspond to a bound state, virtual state or resonance \cite{Badalyan1982}.
The zeros of the Jost function $D(p_1,p_2)$ correspond to the poles of the S-matrix. 
%Since we are interested in a two channel scattering, we consider four Riemann sheets we label as $[tt]$ or the physical sheet, and the unphysical sheets labeled as $[tb]$, $[bb]$, and $[bt]$. 
Analyticity is imposed on \eqref{eq:Sdiag} by requiring
$\lim_{|p_1|,|p_2| \to \infty} D(p_1,p_2) = 1$
for $\text{Im} \ p_n \geq 0$ \cite{Newton, Newton1961}. The Schwarz reflection principle and the hermiticity of the S-matrix below the lowest threshold ensures that for every momentum pole $\bar{p_i}$, there is another pole given by $-\bar{p_i}^*$. All of these must be considered in the construction of the Jost function.

From equation \eqref{eq:Sdiag}, we see that the $S$-matrix is a ratio of two Jost functions. One of the most straightforward ways to construct an analytic, unitary, and symmetric $S$-matrix is by using a Jost function, of the form $D \propto p^{-N}\sum_n^N\alpha_np^n$ where $p=\{p_1,p_2\}$. The extra factor $p^{-N}$ is needed to ensure that as $|p|\rightarrow\infty$ we get $D\rightarrow 1$. The polynomial part, when written in factored form, allows us to form a set of independent zeros of $D(p_1,p_2)$.

\subsection{Independent poles via uniformization scheme}
\label{sec:uniformization}
\hspace{\parindent}
We propose the use of independent poles in the analysis of near-threshold signals for two reasons. First, this is to ensure that the treatment is model-independent. In other parametrizations, such as the Flatté or effective range expansion, fixing one of the poles will necessarily alter the position of the other pole. These parametrizations may be constructed without reference to any models but one can always find an effective coupled-channel potential that can reproduce such pole trajectory \cite{PearceGibson, Frazer1964,Hanhart2014,Hyodo:2014bda}. Second, a parametrization that allows independence of poles can cover a wider model space without violating the expected properties of the S-matrix unlike in other parametrization. This limitation is observed in \cite{Frazer1964} where a specific coupled-channel effective range approximation can only produce poles in either $[bt]$ or $[tb]$ sheet but not in $[bb]$. An S-matrix with independent poles can cover a wider model space without compromising any of the expected properties of S-matrix. 

%The exploration of line shapes near the threshold can be facilitated by controlling the pole positions and their Riemann sheets. We discussed above how to construct an $S$-matrix such that controlling the pole would be convenient. 
It is also important that the amplitude model to be used gives the correct threshold behavior associated with branch point singularity of the two-hadron scattering. The uniformization method introduced in \cite{Kato1965,Newton} and utilized in \cite{Yamada2021,Yamada2020} is an appropriate scheme for our present objective. 
The $k$th channel's momentum in the two-hadron center of mass frame is given by
\begin{equation}
	p_k^2=\dfrac{(s-\epsilon_k^2)
		\left[s-\epsilon_k(\epsilon_k-4\mu_k)\right]}
	{4s}
	\label{eq:actual_p}
\end{equation}
where $\epsilon_k$ is the threshold energy and $\mu_k$ is the reduced mass of the system. The invariant Mandelstam variable $s$ is can be written as
\begin{align}
	s = \epsilon_k^2 + \dfrac{\epsilon_k}{\mu_k}|\vec{p}_k|^2
	\left[1+\mathcal{O}\left(\dfrac{|\vec{p}_k|^2}{\epsilon_k^2}\right)\right]=\epsilon_k^2+q_k^2.
	\label{eq:invariant_s}
\end{align}
where we introduced the new momentum variable $q_k$ for convenience of scaling. 

Instead of constructing the $S$-matrix using the momenta $q_1$ and $q_2$, we introduce the uniformized variable $\omega$ defined by the transformation
\begin{align}
	\omega=\dfrac{q_1+q_2}
	{\sqrt{\epsilon_2^2-\epsilon_1^2}};
		\quad\quad
		\dfrac{1}{\omega}=
		\dfrac{q_1-q_2}{\sqrt{\epsilon_2^2-\epsilon_1^2}}.
			\label{eq:uniformization}	
\end{align}
The linear dependence of the $\omega$ with $(q_1,q_2)$ removes the issue of branch point singularity associated with the threshold. In other words, uniformization reduces the number of complex planes from four energy planes to only one $\omega$ plane. Figure~\ref{1bomegaplane} shows the $\omega$ - plane and a detailed description of such mapping can be seen in \cite{Kato1965,yamada2022near}.  All the relevant halves of the complex energy planes in Fig.~\ref{1aEplane} are on the 1st and 4th quadrants of the $\omega$ plane in Fig.~\ref{1bomegaplane}. %We will discuss in section~\ref{sec:uniformization} how uniformization works and its importance in our method. We now move on to discuss the analytic properties of the $S$-matrix and how to construct it using uniformization.

% \textcolor{red}{We can easily map an energy point from a specific Riemann sheet to the uniformized $\omega$ plane. Initially, let's examine how different Riemann sheets are connected. The [tt] and [bt] sheets are linked along $\epsilon_1<\sqrt{s}<\epsilon_2$. On this energy line, $q_1=\sqrt{s-\epsilon_1^2}$ is a real value, while $q_2=i\sqrt{\epsilon_2^2-s}$ is imaginary. It can be demonstrated that $\omega\omega^*=1$, indicating that energy points within the range $\epsilon_1<\sqrt{s}<\epsilon_2$ are mapped onto a quarter arc of the unit circle in the $\omega$ plane. This arc begins at $\omega=i$ (1st threshold) and ends at $\omega=1$ (2nd threshold). Similar argument can be done for other Riemann sheets that are linked between the two thresholds. Next, the upper half of any complex energy plane is connected to its lower half along the line below the lowest threshold. Here, both $q_1$ and $q_2$ are purely imaginary, which corresponds to purely imaginary $\omega$. That is, the upper half of a complex energy plane is connected to its lower half along the imaginary axis of the $\omega$ plane.}

Referring to equations \eqref{eq:Sdiag} and \eqref{eq:Soffdiag}, we can use a rational Jost function $D(\omega)$ such that the pole of the $S$-matrix is easily determined by its zeros. The simplest Jost function takes the form
\begin{align}
	D(\omega)=\dfrac{1}{\omega^2}
	(\omega-\omega_\text{pole})
	(\omega+\omega_\text{pole}^*)
	(\omega-\omega_\text{reg})
	(\omega+\omega_\text{reg}^*)
	\label{eq:jost}
\end{align}
where we had introduced several factors. The negative conjugate terms are a consequence of the hermiticity of the $S$-matrix below the lowest threshold \cite{Yamada2020,Taylor}. The extra pole $\omega_\text{reg}$, called the regulator, is added to ensure that the $k$th diagonal elements of the S-matrix behave as $S_{kk}(\omega)\rightarrow1$ as $\omega\rightarrow\infty$ \cite{Kato1965,Newton}. In the short-ranged potential scattering theory, we expect the phase shift to vanish at large energies to be consistent with the Levinson's theorem. This expectation can be met when we impose $|\omega_{reg}\omega_\text{pole}|=1$. This means that we have another pole which depend on $\omega_{\text{pole}}$. To ensure that $\omega_{\text{pole}}$ is the only relevant pole that can affect the interpretation of enhancement, we set the regulator to be far from the scattering region. Referring to the uniformized $\omega$ plane in Fig.~\ref{1bomegaplane}, one can minimize the influence of the regulator by placing it either on the $[bb]$ or the $[tb]$ sheet. Note that a regulator pole on the $[bt]$ sheet will result into a structure between the two thresholds, which significantly affects the interpretation of the line shape, hence it is in our best interest to avoid putting a regulator in this region. The simplest regulator we could use following these requirements is $\omega_{\text{reg}} = e^{-i\pi/2}/|\omega_{\text{pole}}|$, where the phase factor ensures that the regulator falls on either the $[tb]$ or $[bb]$ sheet below the lowest threshold.

We reiterate the importance of regulator not affecting the physics of the $S$-matrix. 
%This is assured if the uniformized amplitude only have contributions from nearby poles in the structure seen in the scattering data. 
In \cite{Yamada2020}, the uniformized truncated Mittag-Leffler parametrization neglected the $\omega_\text{reg}$ regulator. Such formulation assumes that the conjugate pole $\omega_{\text{pole}}^*$ is much closer to the scattering region than the $\omega_{\text{reg}}$. The absence of other background poles, especially $\omega_{\text{reg}}$, makes the contribution of $\omega_{\text{pole}}^*$ relevant in the interpretation of the $\Lambda(1405)$, which resulted into a broad line shape in the $\pi\Sigma$ mass distribution. This is the reason why the authors of \cite{Yamada2020} concluded that the $\Lambda(1405)$ requires only one pole in the second Riemann sheet in contrast with the current consensus of two-pole structure interpretation \cite{ParticleDataGroup:2022pth,Wang:2021lth,Meissner:2020khl,Hyodo:2011ur}. Care must be taken in the construction of amplitudes. Removing the contribution of $\omega_{\text{reg}}$, or any possible background poles, as emphasized in \cite{Kato1965} might lead into misinterpretation of line shapes.

In other parametrizations, such as the K-matrix model (see for example \cite{Kuang2020}), the equation for the poles are typically quartic in the channel momenta. One may set the Riemann sheet of the desired relevant pole by adjusting the coupling parameters but there is always a tendency that a shadow pole will appear on the physical sheet \cite{SombilloPhysRevD.104.036001,Eden1964shadow}. The linear dependence of the uniformized variable $\omega$ on the channel momenta in Eq.\eqref{eq:uniformization} guarantees that no shadow pole is produced using the Jost function in \eqref{eq:jost}. The regulator pole, which can be controlled in our formulation, ensured that the S-matrix will not violate causality. In situation where there is an actual shadow pole, an independent pole can be added to the Jost function with no direct relation to the main pole. That is, one can freely place the pole and its shadow in any position without being restricted by some coupling parameter. With all of these considerations, the full Jost function with different combinations of $N$ independent poles takes the form:
\begin{equation}
	D(\omega)=\dfrac{1}{\omega^{2N}}
	\prod_{\{\text{pole, reg}\}}^{N}
	(\omega-\omega_\text{pole})
	(\omega+\omega_\text{pole}^*)
	(\omega-\omega_\text{reg})
	(\omega+\omega_\text{reg}^*).
	\label{eq:bigjost}
\end{equation}
Using the independent-pole form of the Jost function in Eq.\eqref{eq:bigjost}, the two-channel S-matrix elements satisfying unitarity, analyticity, and hermiticity below the lowest threshold takes the form
\begin{align}\label{eq:smatrix}
			S_{11}(\omega)=\dfrac{D(-1/\omega)}{D(\omega)}; \quad
			S_{22}(\omega)=\dfrac{D(1/\omega)}{D(\omega)}; \quad
			\text{det}(S)=\dfrac{D(-\omega)}{D(\omega)}. 
\end{align}
The scattering amplitude $T_{j,k}(\omega)$ can be obtained from the relation $S_{j,k}=\delta_{j,k}-2iT_{j,k}$ where $\delta_{j,k}$ is the Kronecker delta.

\section{Ambiguous line shapes}\label{sec:III}
The independent poles of the full S-matrix are determined by the zeros of the $D(\omega)$. It is possible that the zero of denominator cancels the zero of the numerator for one of the S-matrix element, say $S_{11}$. This means that such pole will not manifest in the line shape as if it did not exist at all. 
%The independence of the poles from each other opens up the possibility of the simultaneous zeros of the numerator and denominator. This leads to no observable contribution to the scattering cross section, at least when probing only one elastic channel. 
This subtle features of the S-matrix is important as we can potentially miss out other possible pole configurations if we probe only one element of the full $S$-matrix. In this section, we first present on how this ambiguity arises. 
We emphasize that this ambiguity is dependent on the parametrization of the $S$-matrix. 
In particular, the ambiguity of the formalism we use arises from the parametrization of the regulator. As an application, we discuss the implication of the ambiguous line shapes in the context of the $P_\psi^N(4312)^+$ signal. We propose that such ambiguity can be removed by probing the off diagonal term of the $S$-matrix. We close the section by discussing the importance of probing the off diagonal $S$-matrix terms and the limitation of the effective range expansion.

\subsection{Emergence of ambiguity}
Here, we point out that there exists a pole configuration that has no effect on one of the elements of the full S-matrix. We start by looking at the pole configuration of one of the $S$-matrix element that is equal to unity. We focus on the ambiguity of the $S_{11}$ channel. Recall that we construct the $S_{11}(\omega)$ element as
\begin{align}\label{eqn:S11explicit}
\begin{split}
S_{11}(\omega) &= \frac{D(-1/\omega)}{D(\omega)} \\& = \omega^{4N}\frac{\prod_{\{\text{zero,reg}\}}^N(-1/\omega-\omega_{\text{zero}})(-1/\omega-\omega_{\text{reg}})(-1/\omega+\omega_{\text{zero}}^*)(-1/\omega+\omega_{\text{reg}}^*)}{\prod_{\{\text{pole,reg}\}}^N(\omega-\omega_{\text{pole}})(\omega-\omega_{\text{reg}})(\omega+\omega_{\text{pole}}^*)(\omega+\omega_{\text{reg}}^*)}\\& = \omega^{4N}\frac{\prod_{\{\text{zero,reg}\}}^N(-1/\omega-\omega_{\text{zero}})\times(\text{reg.})\times (-\text{c.c})}{\prod_{\{\text{pole,reg}\}}^N(\omega-\omega_{\text{pole}})\times(\text{reg.})\times (-\text{c.c})} 
\end{split} 
\end{align}
where the notation in the last line is introduced for convenience. The terms reg. and c.c. denote the regulator and complex conjugate of the zeros (or poles) of the preceding terms, respectively. With some foresight, we consider the pair of poles $\omega_{\text{pole 1}} = a$ and $\omega_{\text{pole 2}}= -1/a$. From \eqref{eqn:S11explicit}, the $S_{11}(\omega)$ matrix element reads as
\begin{align}
\begin{split}
S_{11}(\omega) &= \omega^{8}\times\frac{(-1/\omega - a)}{(\omega-a)}\cdot \frac{(-1/\omega +1/a)}{(\omega+1/a)}\times (\text{reg}) \times (-\text{c.c.}) \\&= \omega^{8} \times \left(-\frac{1}{\omega^2}\right) \times \left(-\frac{1}{\omega^2}\right) \times \left(\frac{1}{\omega^4}\right) \\&= 1
\end{split}
\end{align}
The pair of poles $\omega_{\text{pole 1}} = a$ and $\omega_{\text{pole 2}}= -1/a$ entails a unitary $S$-matrix. We will call such combination of poles that lead to a unitary $S$-matrix element as ``ambiguous pair poles." In terms of momentum, recalling equation \eqref{eq:uniformization}, the pair poles $\omega_{\text{pole 1}} = a$ and $\omega_{\text{pole 2}} =-1/a$ translates into
\begin{align}\label{eqn:SeeThis}
\frac{q_1^{(2)} + q_2^{(2)}}{\sqrt{\epsilon_2^2 - \epsilon_1^2}} = \frac{q_2^{(1)} - q_1^{(1)}}{\sqrt{\epsilon_2^2 - \epsilon_1^2}}
\end{align}
where $q_n^{(m)}$ denotes the momentum of the $n$th channel of the $m$th pole. It follows from \eqref{eqn:SeeThis} that $q_1^{(2)} + q_1^{(1)} = 0$ for the equation to hold; the takeaway is that the imaginary part of the $q_1$ pair must be opposite. This implies that the pair poles should either be found separately on the $[tt]$ and $[bt]$ sheet \textit{or} on the $[bb]$ and $[tb]$ sheets. The former instance is prohibited since it will violate analyticity and causality. Hence, we consider two poles in which one is located at the $[tb]$ sheet and the other at the $[bb]$ sheet an ambiguous pair pole if $\omega_{\text{pole}[tb]}\omega_{\text{pole}[bb]} = -1$. Moreover, their regulators are given by $\omega_{\text{reg of $[bb]$ pole}} = -ia$ and $\omega_{\text{reg of $[tb]$ pole}} = i/a$, located on the $[tb]$ and $[bb]$ sheets, respectively.

The condition for the ambiguous pair poles states no specific magnitude for it to occur, only that they must depend on each other as $\omega_{\text{pole}[tb]}\omega_{\text{pole}[bb]} = -1$. Despite that these ambiguous pair poles have no effective contribution in the $T_{11}$ line shape, they cannot be ignored especially when they are near the threshold. If we probe either the $T_{12}$ or $T_{22}$ channel, we will observe a difference between the two configurations. We then assert that one must be rigorous in using certain approximations. For example, without proper justification, the prescription $\mathrm{d}N/\mathrm{d}\sqrt{s} \propto |T_{11}|^2$ might potentially miss out pole configurations with ambiguous pair poles.

Through a similar argument, we could assert that the same conclusion holds for the $S_{22}$ channel. The ambiguous pair for the $S_{22}$ channel should lie on the $[bb]$ and $[bt]$ sheet and satisfies $\omega_{\text{pole}[bt]}\omega_{\text{pole}[bb]} = 1$. However, this will only be possible if we reassign the regulator of the pole at the $[bt]$ sheet as $\omega_{\text{reg of $[bt]$}} = e^{i\pi/2}/|\omega_{\text{pole in $[bt]$}}|$, i.e., the regulator should be placed on the $[tt]$ sheet instead of the $[bb]$ sheet. We consider for a moment this pole on the $[bt]$ sheet and its regulator. Recalling equations \eqref{eq:bigjost} and \eqref{eq:smatrix}, we have
\begin{align}
S_{22} = \omega^4 \frac{(1/\omega-\omega_{[bt]})(1/\omega+\omega_{[bt]}^*) }{(\omega-\omega_{[bt]})(\omega+\omega_{[bt]}^*)}  \cdot \frac{(1/\omega-\omega_{[tt]})(1/\omega+\omega_{[tt]}^*)}{(\omega-\omega_{[tt]})(\omega+\omega_{[tt]}^*)}
\end{align}
where the second fraction are the regulator terms and its complex conjugates. Since $\omega_{tt}$ is located at the imaginary axis, the denominator of the second fraction is equal to $(\omega-\omega_{[tt]})^2$, which makes the singularity a double pole. We have to recall that state correspondence is always a simple pole \cite{Newton, Taylor} and hence, our regulator assignment on the $[tt]$ sheet does not invalidate causality. Moreover, this 2nd order $[tt]$ pole is faraway from the threshold, and hence its presence is not relevant in the interpretation of amplitude line shape.

\subsection{Implication of the ambiguous line shapes}
To demonstrate the ambiguous line shape and its consequences, we look into the LHCb $P^N_\psi(4312)^+$ signal. We reconstruct its scattering amplitude using the data and best fit found in \cite{Pc4312}. To facilitate the discussion of this section, we will be using the pole counting method by Morgan \cite{Morgan1992} to analyze the resulting scattering amplitudes. The idea hinges on the fact that non-potential resonances occurring very close to an $s$-wave threshold is associated with poles on the $[bt]$ and $[bb]$ sheets of the energy plane \cite{MorganPennington1991, morgan1993decay}. Simply put, the pole counting method states that a single pole on the $[bt]$ sheet indicates a predominantly molecular bound state whereas poles appearing on the $[bt]$ and $[bb]$ sheets both close to the threshold indicates a state dominated by its compact component. 
%A more detailed approach will be done in a different work.

\begin{figure}[h!]
        \centering
        \begin{subfigure}[b]{0.45\columnwidth}
            \centering
            \includegraphics[scale = 0.25]{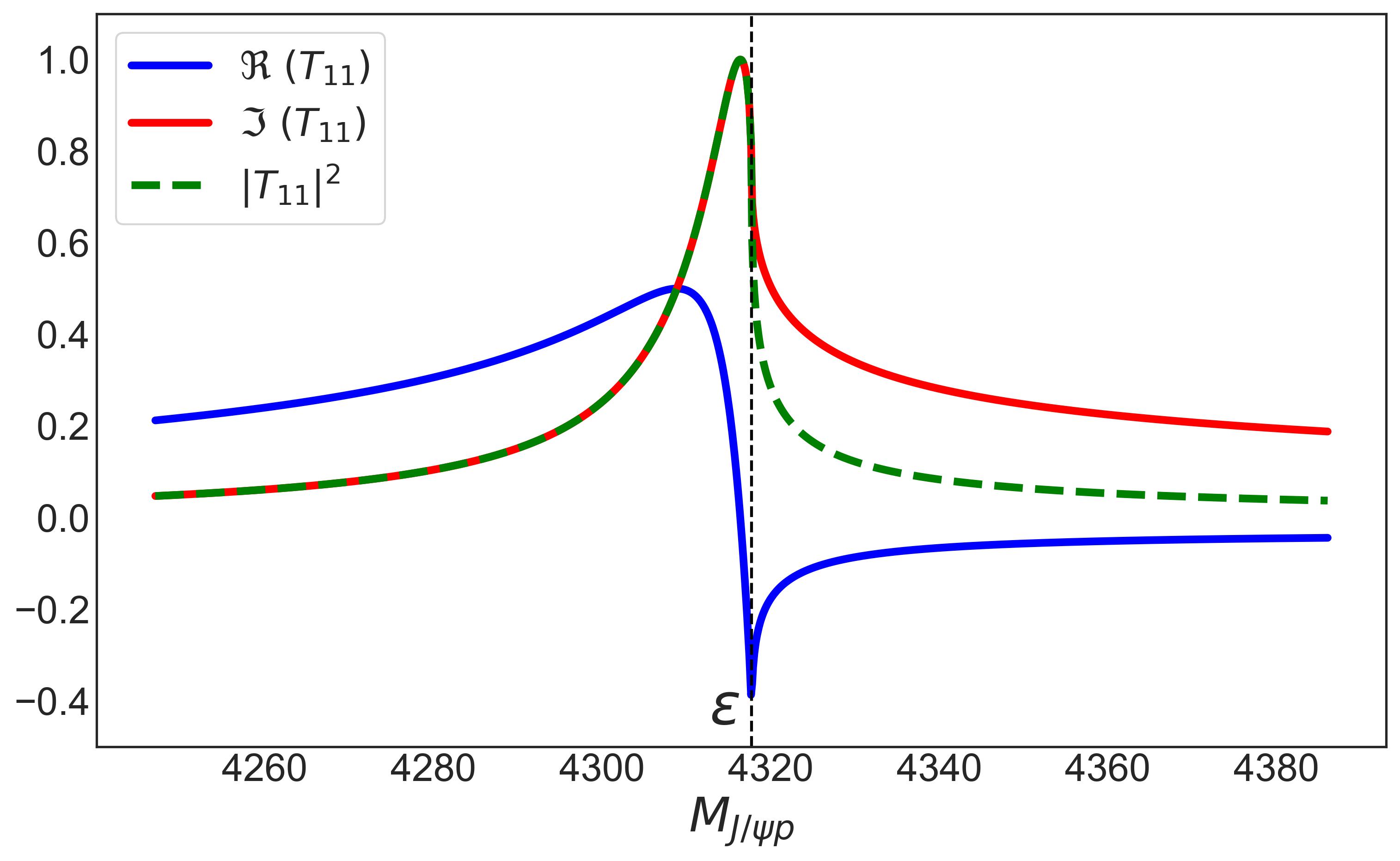}
            \caption[]
            % {{\small Displacing Re([bb]) while fixing the [tb] pole. The displacement below the original mass accumulates as shown by the blue lines.}}    
            
        \end{subfigure}
        %\hfill
        \begin{subfigure}[b]{0.45\columnwidth}  
            \centering 
            \includegraphics[scale = 0.25]{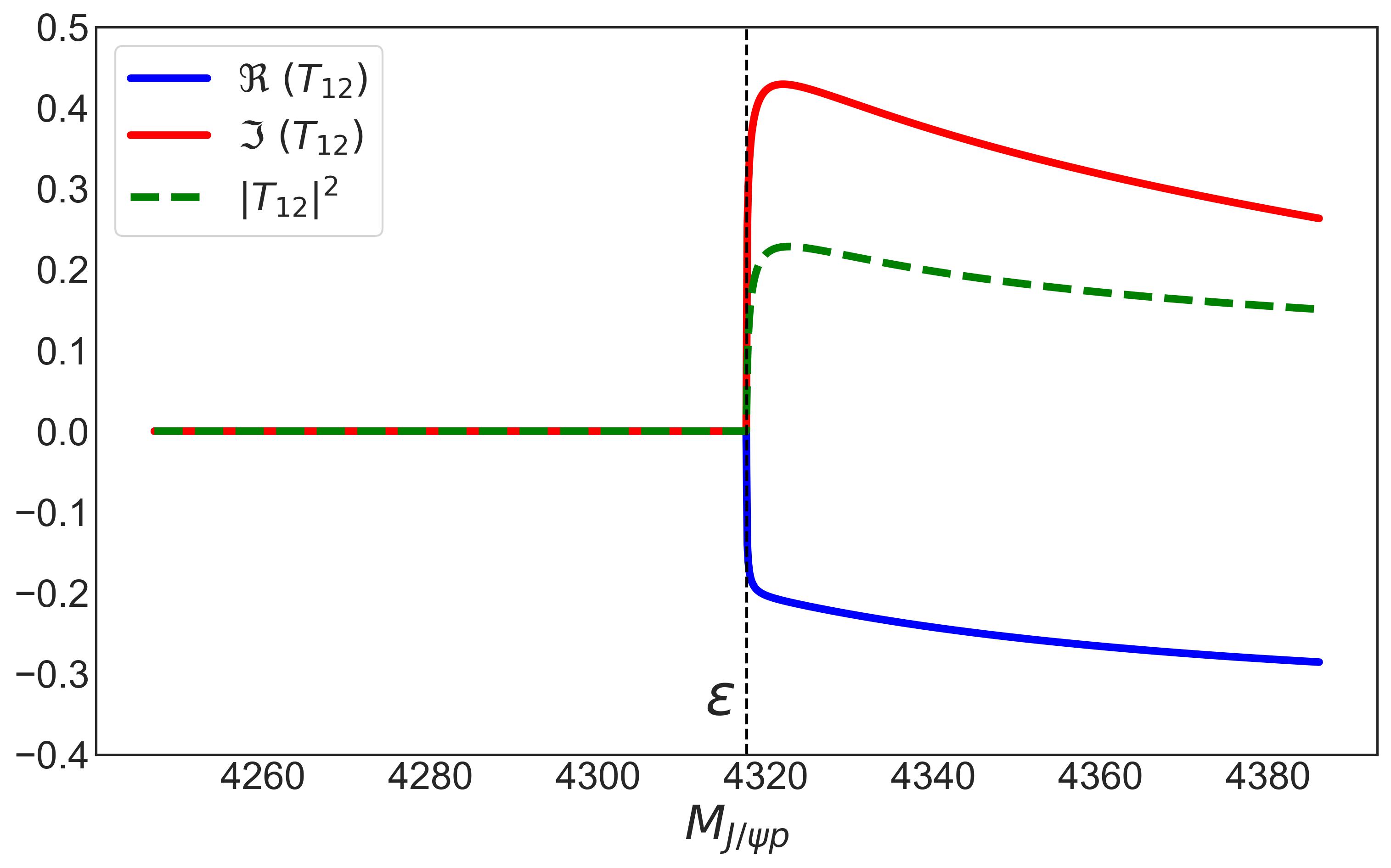}
            \caption[]
            % {{\small Displaced Im([bb]) while fixing the [tb] pole. The displacement below the original width accumulates as shown by the blue lines.}}    
            
        \end{subfigure}
        \vskip\baselineskip
        \begin{subfigure}[b]{0.45\columnwidth}   
            \centering 
            \includegraphics[scale = 0.2]{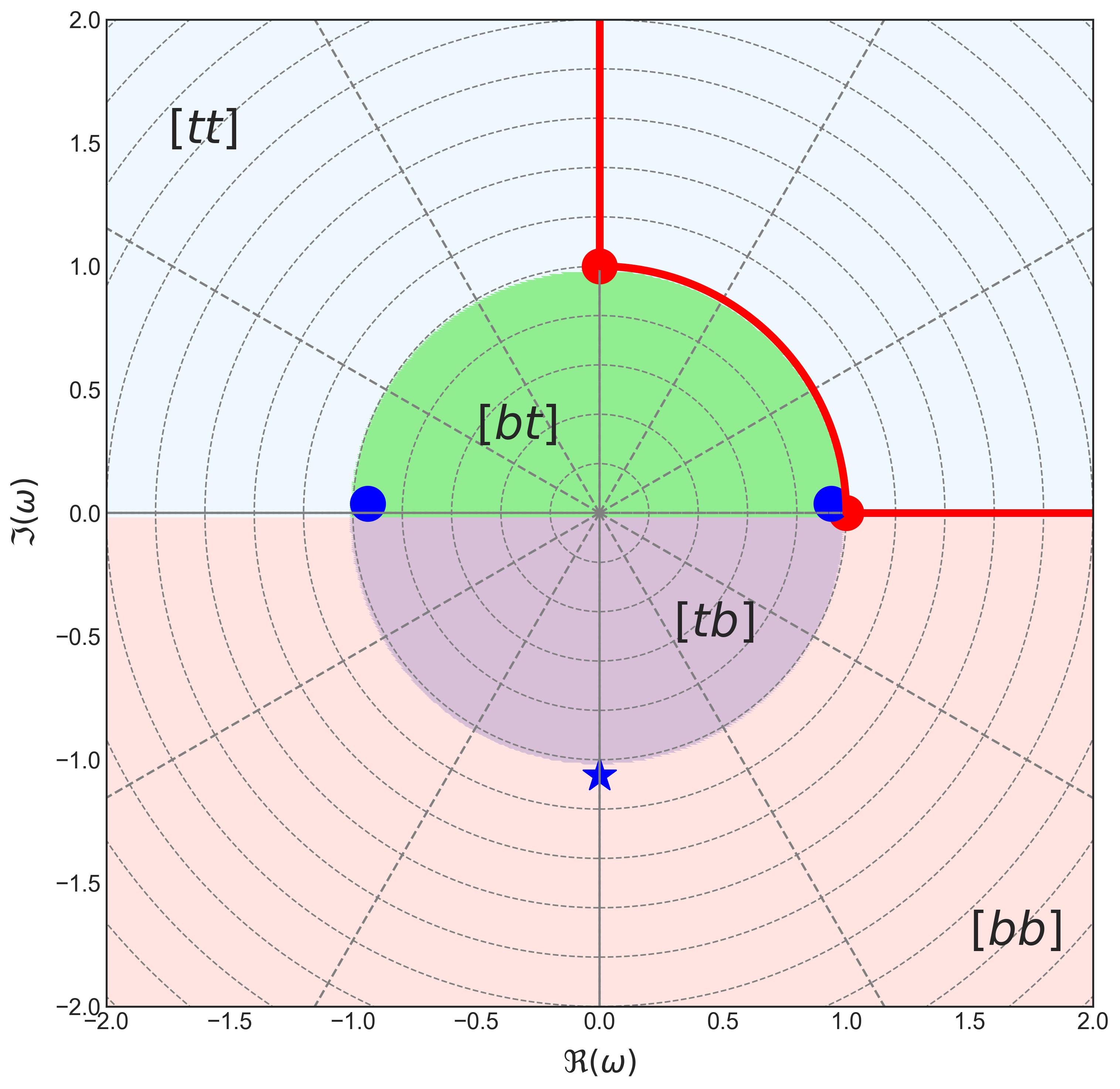}
            \caption[]
            % {{\small Displaced Re([tb]) while fixing the [bb] pole. The displacement below the original mass accumulates as shown by the blue lines.}}    
            
        \end{subfigure}
        %\hfill
        \begin{subfigure}[b]{0.45\columnwidth}   
            \centering 
            \includegraphics[scale = 0.25]{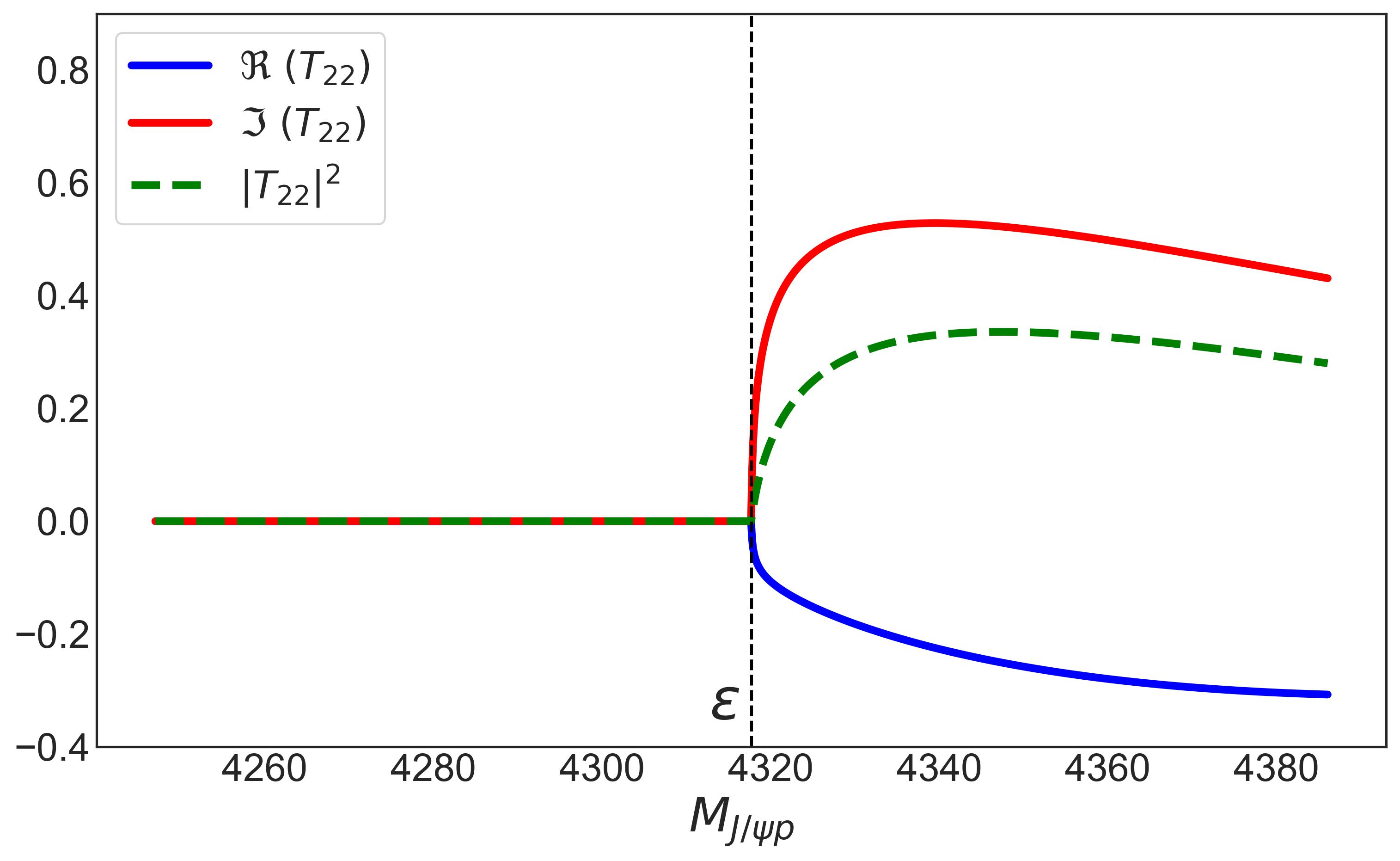}
            \caption[]
            % {{\small Displaced Im([tb]) while fixing the [bb] pole. The displacement below the original width accumulates as shown by the blue lines.}}    
            
        \end{subfigure}
        \caption[]
        {Elements of the $T$-matrix with a pole on the $[bt]$ sheet, signifying a molecular nature based on pole counting method.}
        \label{fig:bt}
    \end{figure}

First, we consider a $1$-pole configuration with a background pole. We placed the main pole on the $[bt]$ with real part $m = 4319.8$ MeV and width $\Gamma = 9.2$ MeV. The background pole is added on the $[tb]$ sheet, below the first channel threshold $3184.9$ MeV with width $\Gamma = 1000$ MeV. Figure~\ref{fig:bt} shows the $T$-matrix elements of this $1$-pole configuration.

By looking only at the $T_{11}$ matrix element, its sharp peak near the $\Sigma_c\bar{D}$ threshold hints a signature of a molecular state. We can back this up by appealing to the pole counting method which, for an isolated pole in $[bt]$ sheet, suggests a molecular nature of the observed  signal. On the other hand, both the bottom-top approach done in \cite{Pc4312} and \cite{DeepLearningExHad} which utilized the form
\begin{align}
\frac{\mathrm{d}N}{\mathrm{d}\sqrt{s}} = \rho(s) \left[|P(s)T(s)|^2 + B(s)\right]
\end{align}
where $P(s)$ and $B(s)$ are smooth functions, $\rho(s)$ the three-body phase space and $T(s)$ being the $T_{11}$ matrix element had concluded that the signal is likely due to a virtual state. As mentioned earlier, near-threshold virtual states are molecular in nature as long as range corrections can be neglected \cite{matuschek2021nature}, and hence the agreement with the pole couting method and the conclusions in \cite{Pc4312} and \cite{DeepLearningExHad}. However, it is noteworthy to mention that the recent investigation of the same signal using deep learning in \cite{ZHANG2023981} favors the compact interpretation. 

% It is noteworthy to mention that the parametrization
% \begin{align}
% \frac{\mathrm{d}N}{\mathrm{d}\sqrt{s}} = \rho(s) \left[|P(s)T(s)|^2 + B(s)\right]
% \end{align}
% where $P(s)$ and $B(s)$ are smooth functions, $\rho(s)$ the three-body phase space and $T(s)$ being the $T$-matrix was used in \cite{Pc4312}, \cite{DeepLearningExHad}. Both studies only used the elastic $T_{11}$ channel in their parametrization. 

\begin{figure}[h!]
        \centering
        \begin{subfigure}[b]{0.45\columnwidth}
            \centering
            \includegraphics[scale = 0.25]{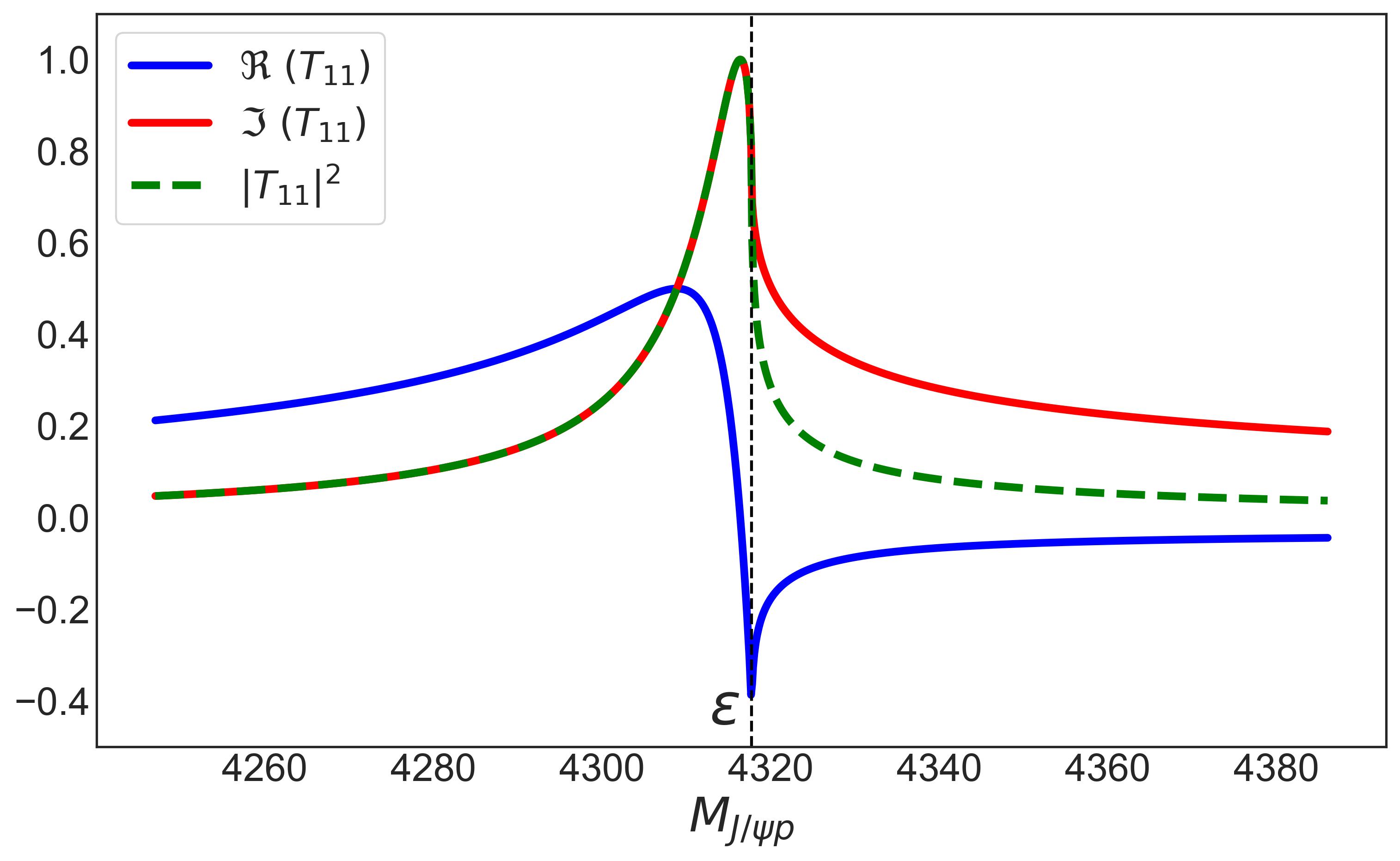}
            \caption[]
            % {{\small Displacing Re([bb]) while fixing the [tb] pole. The displacement below the original mass accumulates as shown by the blue lines.}}    
            
        \end{subfigure}
        %\hfill
        \begin{subfigure}[b]{0.45\columnwidth}  
            \centering 
            \includegraphics[scale = 0.25]{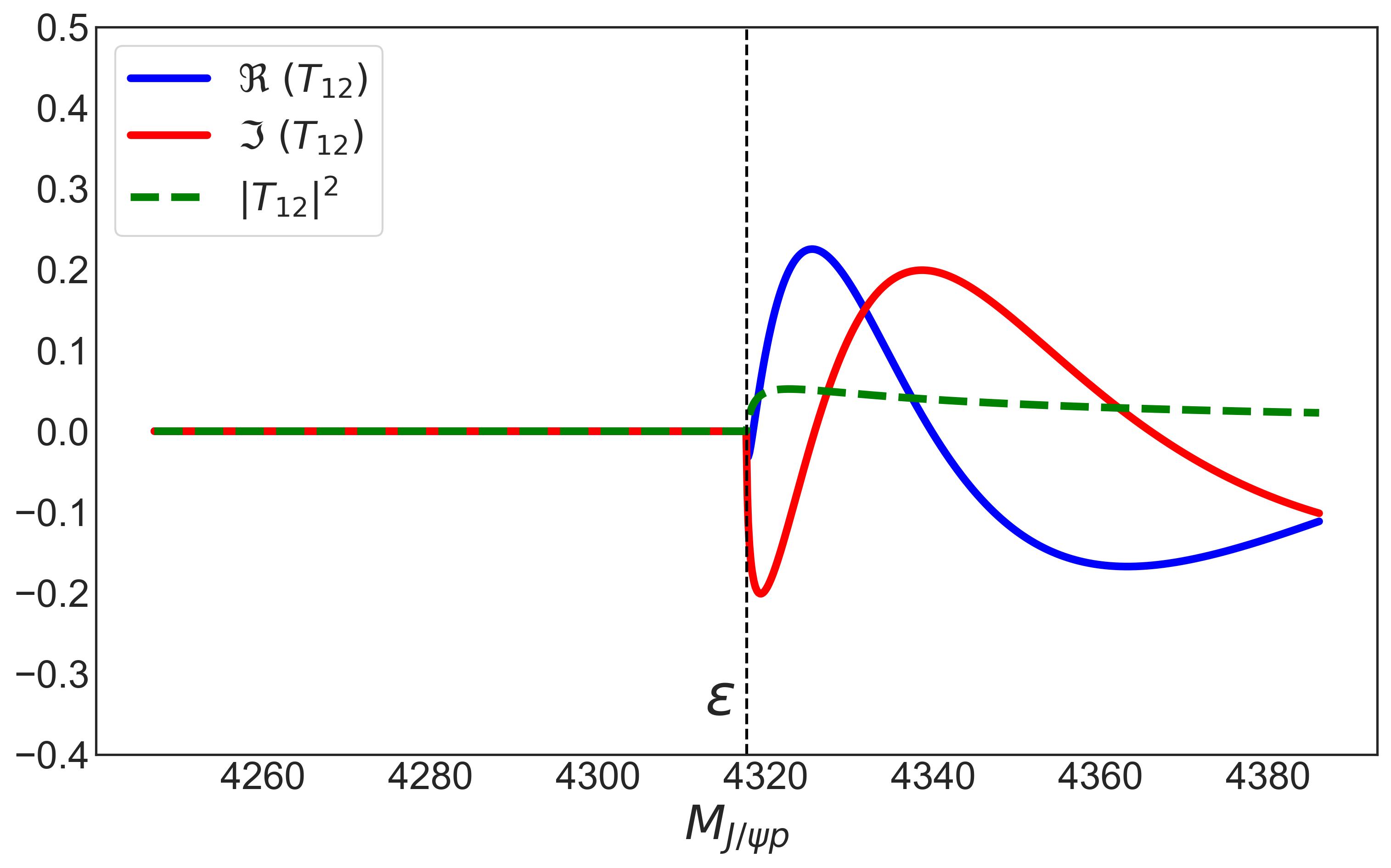}
            \caption[]
            % {{\small Displaced Im([bb]) while fixing the [tb] pole. The displacement below the original width accumulates as shown by the blue lines.}}    
            
        \end{subfigure}
        \vskip\baselineskip
        \begin{subfigure}[b]{0.45\columnwidth}   
            \centering 
            \includegraphics[scale = 0.2]{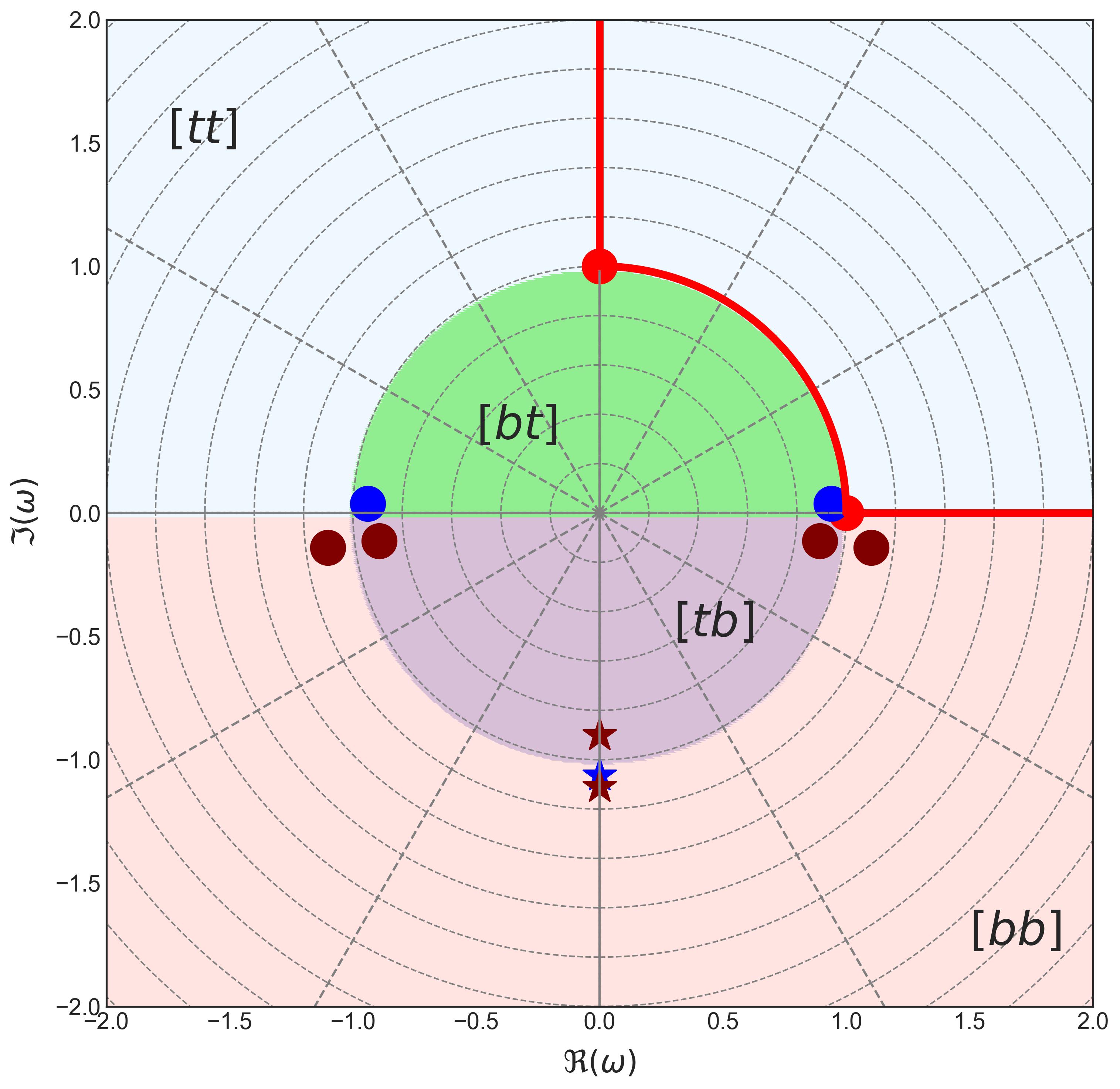}
            \caption[]
            % {{\small Displaced Re([tb]) while fixing the [bb] pole. The displacement below the original mass accumulates as shown by the blue lines.}}    
            
        \end{subfigure}
        %\hfill
        \begin{subfigure}[b]{0.45\columnwidth}   
            \centering 
            \includegraphics[scale = 0.25]{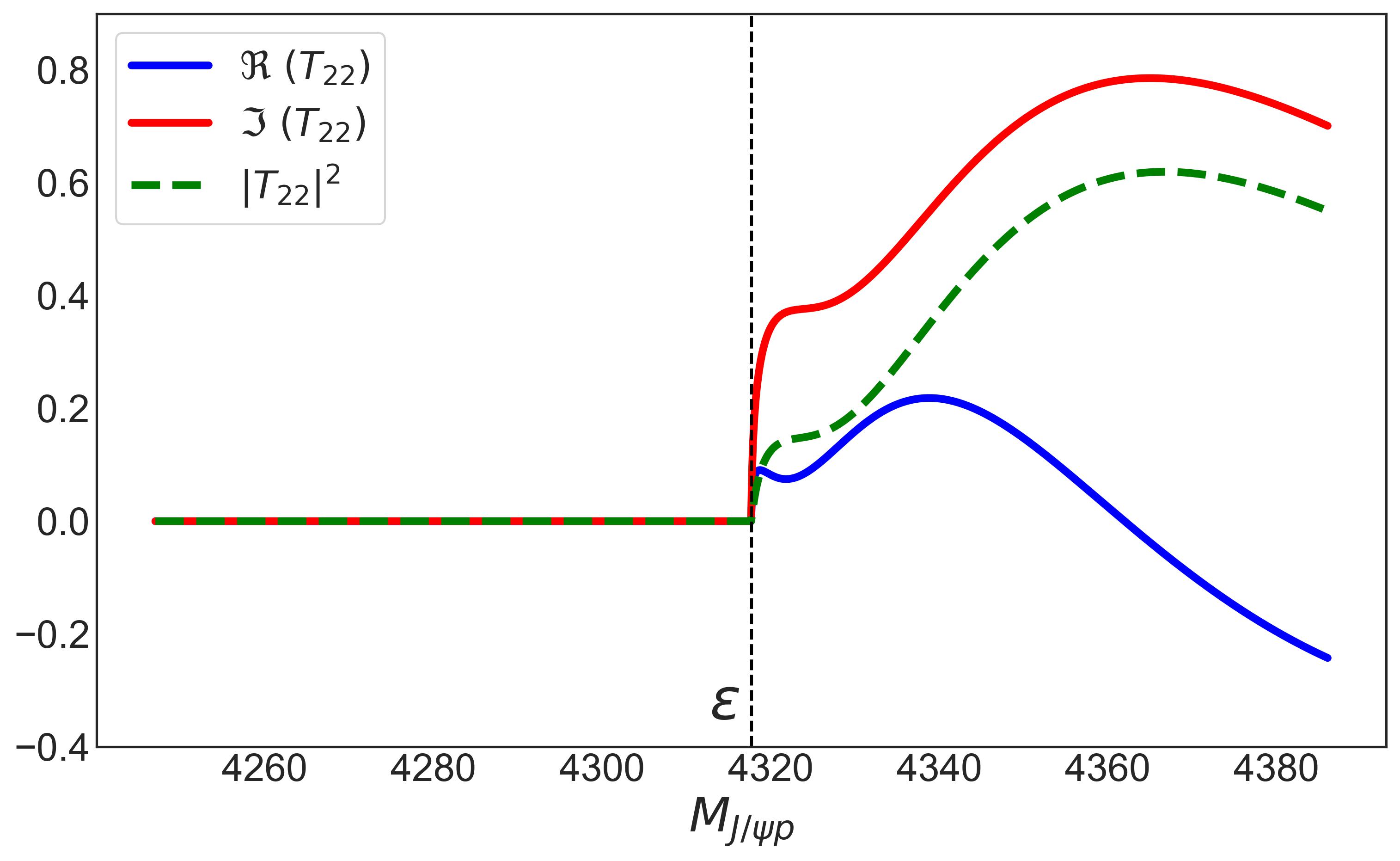}
            \caption[]
            % {{\small Displaced Im([tb]) while fixing the [bb] pole. The displacement below the original width accumulates as shown by the blue lines.}}    
            
        \end{subfigure}
        \caption[]
        {Elements of the $T$-matrix with an isolated pole on the $[bt]$, $[bb]$, and $[tb]$ sheets. Can be interpreted having a non-molecular nature based on pole counting method.}
        \label{fig:bt_amb}
    \end{figure}

We move on to the next configuration. We use the same isolated pole in $[bt]$ sheet from the previous configuration plus an additional ambiguous pair poles on the $[bb]$ and $[tb]$ sheets. The values of the added poles are set to produce an ambiguous pair of poles as described in the previous subsection. Specifically, we set the pair poles such that their real parts are $4317.73$ MeV and $12.5$ MeV for its widths. The three-pole configuration and its $T$-matrix elements are shown in Fig.~\ref{fig:bt_amb}.
The interpretation of 3-pole configuration is now outside the direct applicability of the pole counting argument. We are confronted with different possible scenarios for the occurrences of such pole structures. One naive interpretation is that the $[bt]$ sheet pole is a molecular state of the $\Sigma_c\bar{D}$ channel with the added $[bb]$ and $[tb]$ poles as some non-molecular state that is strongly coupled to $\Sigma_c\bar{D}$ channel. However, such interpretation is unlikely since the real parts of the $[bb]$ and $[tb]$ poles are exactly placed at the second threshold. This means that there is no available phase space for the unstable state to decay into the second channel. 

A more plausible interpretation is that the $[tb]$ sheet pole is a virtual state of the $\Sigma_c\bar{D}$ channel and the remaining $[bt]$ and $[bb]$ poles correspond to a compact state that is strongly coupled to the $J/\psi p$ channel. Unlike the first interpretation, there is enough phase space for such a compact resonance to decay into $J/\psi p$ channel. Interestingly, this scenario matches the hybrid model proposed in \cite{Yamaguchi:2017zmn,Yamaguchu2020}. Comparing the $T_{11}$ line shape in Fig.~\ref{fig:bt}, which admits a purely molecular interpretation, and $T_{11}$ in Fig.~\ref{fig:bt_amb}, one can interpret that a compact state combined with a virtual state can lead to a purely molecular-like interpretation. That is, the compact state enhances the attraction of the hadrons in the higher mass channel.

Certainly, there is an obvious ambiguity in the interpretation of  $T_{11}$ line shape with the 1-pole and the 3-pole structures. %Given a suitable potential, the a 1-pole configuration and a 3-pole configuration can be reproduced. 
%The pole counting method states that a pole on the $[bt]$ and $[bb]$ sheet implies the non-molecular nature of the state. In the pole configuration we consider in Figure~\ref{fig:bt_amb}, we have three poles and the pole counting method cannot be used hastily. For this study's purposes, we proceed using the pole counting method by arguing that the ``excess" pole on the $[tb]$ sheet came upon from the interaction of the poles on the $[bt]$ and $[bb]$ sheets. 
%Clearly, rigor demands the dynamics of the argument. We can do this by designing an appropriate potential and this will be done in a separate work. 
 %Comparing the $T_{11}$ line shape of figure~\ref{fig:bt_amb} with the $T_{11}$ line shape of figure~\ref{fig:bt}, we see that they  are identical. 
 Their distinction will only become evident when we probe either the off-diagonal $T_{21}$ element or the elastic $T_{22}$ amplitude. Considering that the available data came from the decay $\Lambda_b\rightarrow K^{-}J/\psi p$, it is appropriate to include the contribution of $T_{21}$ transition when computing for the invariant mass distribution of $J/\psi p$ to determine the presence of ambiguous pair of poles.

% This is especially important, given that there are a number of interpretations of the $P_\psi^N(4312)^+$ signal (see e.g. \cite{KinematicsPhysRevLett.124.072001}).

% \cite{DeepLearningExHad} - pilloni, virtual state 2022 deeplearning

% \cite{KinematicsPhysRevLett.124.072001} - 2020 coupled channel formalism - molecular picture; is a $\Sigma_c\bar{D}$ bound state with $J^P = 1/2^-$

% \cite{chengliumolecular} - 2019 isospin breaking effects to get mass, rearrangement decay properties. Both the masses and decay properties of $P_\psi^N(4312)$ can be understood if one treats it as $J^P = 3/2^-$ compact state or $J^P = 1/2^-$ $S$-wave state. The molecular model with short range contributions and coupled channel effects is more realistic.

% \cite{QCDsumrules} - 2019 QCD sum rules - molecular picture; is the $[\Sigma_c^{++}\bar{D}^-]$ bound state with $J^P = 1/2^-$

% \cite{ALI2019365compact} - 2019 In here, the Pc4312 resonance was interpreted in the compact diquark model as hidden-charm diquark-diquark-antiquark baryon with $J^P=3/2^+$.

\subsection{On the necessity of probing the $T_{12}$ amplitude}
The condition for the existence of ambiguous pair poles, i.e. poles in different Riemann sheets but with exactly the same position, is highly fine tuned. More realistic situation may have poles in different Riemann sheets but not necessarily with the same position due to possible contamination of coupled channel effects. That is, there maybe a slight difference in the line shape of $T_{11}$ for different pole structures. However, when the error bars of the experimental data are large then any distinction among the $T_{11}$s with different pole structures will no longer be useful. The inclusion of $|T_{21}|^2$ might help since the 1-pole structure has slightly larger $|T_{21}|^2$ in comparison with the 3-pole structure. 
%encompass the ambiguity near the threshold energy, then the inclusion of the off-diagonal $S_{12}$ term is important in the interpretation of overall scattering cross section. We provide a qualitative observation below.
\begin{figure}[t!]
	\centering
	\begin{subfigure}[b]{0.45\columnwidth}
		\centering
		\includegraphics[scale = 0.25]{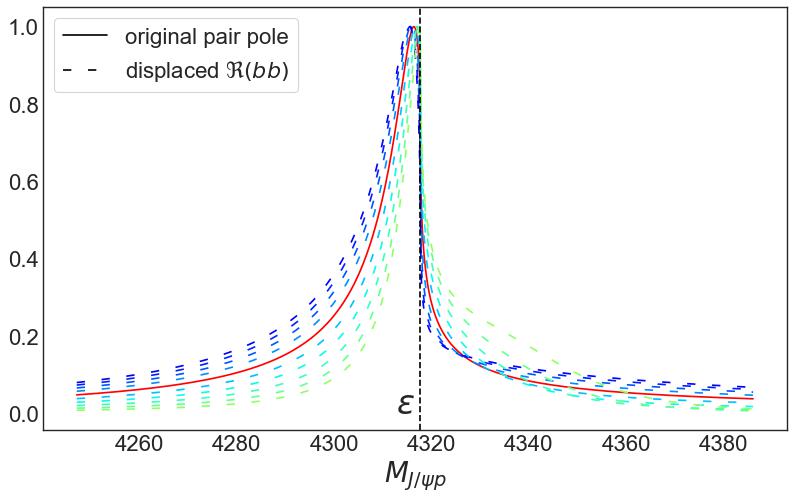}
		\caption[]
		% {{\small Displacing Re([bb]) while fixing the [tb] pole. The displacement below the original mass accumulates as shown by the blue lines.}}    
		
	\end{subfigure}
	%\hfill
	\begin{subfigure}[b]{0.45\columnwidth}  
		\centering 
		\includegraphics[scale = 0.25]{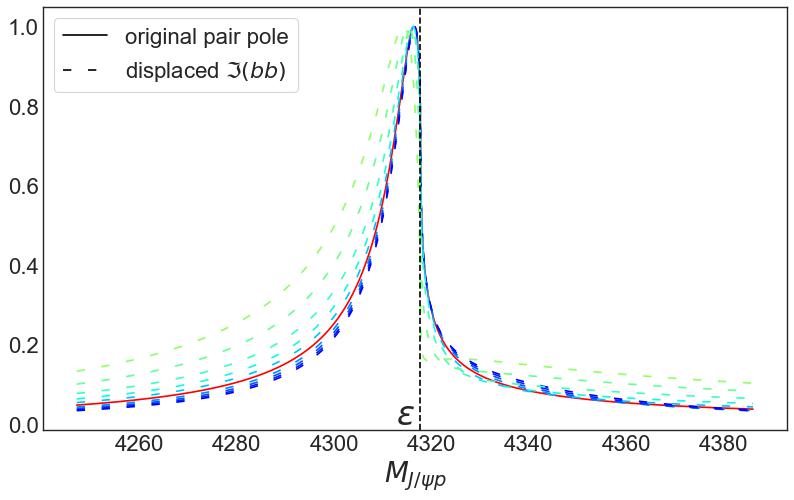}
		\caption[]
		% {{\small Displaced Im([bb]) while fixing the [tb] pole. The displacement below the original width accumulates as shown by the blue lines.}}    
		
	\end{subfigure}
	\vskip\baselineskip
	\begin{subfigure}[b]{0.45\columnwidth}   
		\centering 
		\includegraphics[scale = 0.25]{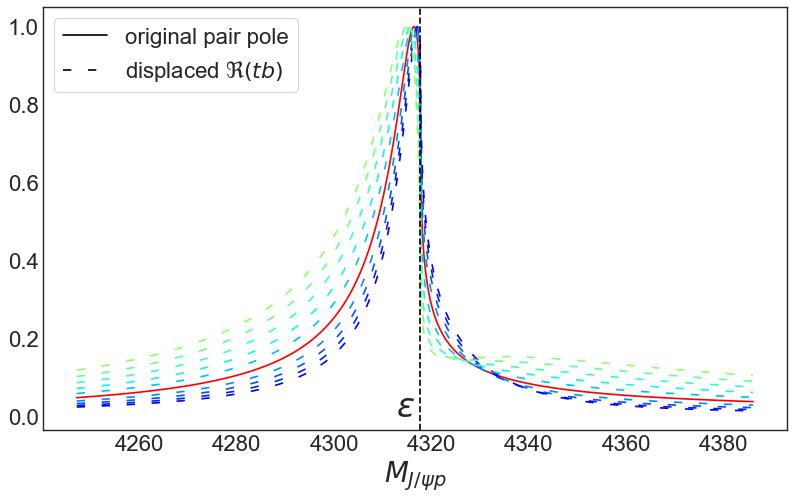}
		\caption[]
		% {{\small Displaced Re([tb]) while fixing the [bb] pole. The displacement below the original mass accumulates as shown by the blue lines.}}    
		
	\end{subfigure}
	%\hfill
	\begin{subfigure}[b]{0.45\columnwidth}   
		\centering 
		\includegraphics[scale = 0.25]{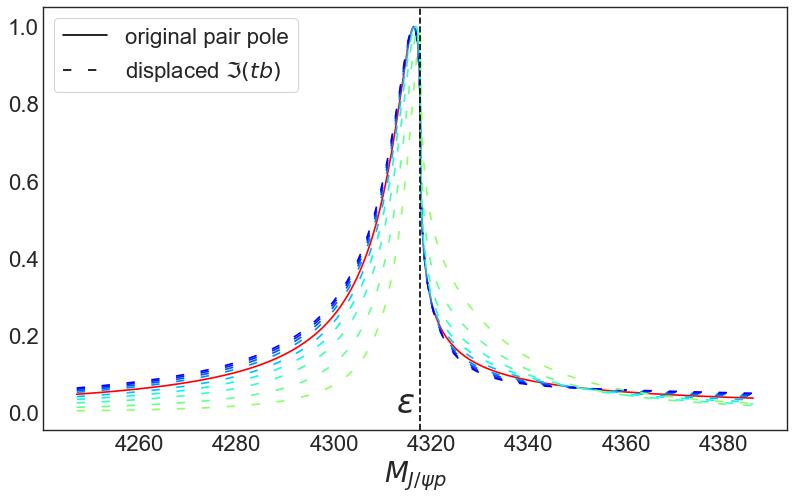}
		\caption[]
		% {{\small Displaced Im([tb]) while fixing the [bb] pole. The displacement below the original width accumulates as shown by the blue lines.}}    
		
	\end{subfigure}
	\caption[]
	{Displacing one of the pair poles and fixing the other. The darker color hues represent the displacement below the original $\Re$ ($\Im$) and the lighter hues represent the displacement above the original $\Re$ ($\Im$) values. Quantitatively, there is little to no difference near the threshold.}
	\label{fig:pole displacement fx}
\end{figure}

To further demonstrate the importance of $|T_{21}|^2$, we consider the $P^N_\psi(4312)^+$ in the invariant mass spectrum of $J/\psi p$. We construct an $S$-matrix similar to that of the 3-pole configuration example above. Afterwards, we displace either the real or the imaginary part of one of the pair poles while fixing in place the other. In this way, the ambiguous pair poles are less fine tuned. The respective $T_{11}$ amplitude line shapes are then plotted in figure~\ref{fig:pole displacement fx}. Notice
that the tail of line shape is sensitive to the presence of ambiguous poles. However, the peak structure for all cases are almost similar despite their differences in their tail. This might entail a problem if the error bars obscured the distinction of ambiguous line shapes near the threshold. It is also possible that the line shape of the slightly displaced ambiguous pair poles might be absorbed by the background parametrizations if the amplitude ansatz is limited to produce one isolated pole near the threshold. It was shown in \cite{Frazer1964} that the effective range expansion cannot produce a pole on the $[bb]$ sheet due to the impossibility of making the real and imaginary parts of the amplitude’s denominator simultaneously equal to zero. Such restriction limits the model space of a given line shape to molecular-like bound or virtual states and immediately rules out compact state interpretation.

\section{Conclusion and outlook}\label{sec:V}
The method of independent S-matrix poles is a useful tool in analyzing near-threshold phenomena in a model-independent way. We have shown that in our formulation, it is possible that some of the commonly used analyses missed out important physics in probing the nature of near-threshold enhancements. The parametrized background may absorb the relevant physics if the parametrization used can only cover a limited model space. Other elements of the full S-matrix can be useful in providing a more rigorous interpretations of the observed signals.

We used the independent S-matrix poles formulation to study the $P_\psi^N(4312)^+$ in the invariant mass spectrum of $J/\psi p$. It turned out that, by focusing only on the contribution of $T_{11}$ in the overall line shape of the distribution, one cannot rule out yet the compact pentaquark interpretation. Our result gives credence to the deep learning analysis made in \cite{ZHANG2023981} and the improved data in \cite{GlueX:2023pev} together with its corresponding analysis in \cite{Strakovsky:2023kqu}. It is also worth noting that a three-channel analysis upholding the principles of S-matrix in \cite{Winney:2023oqc} shows evidence of poles near the scattering region, emphasizing the importance of strong coupling with higher channels. Indeed, sophisticated methods catering all possibilities must be considered to obtain a definitive interpretation of the pentaquark signals.

%We have shown that scattering amplitudes may posses two different pole configurations by the virtue of ambiguous pair poles. Its effects are felt strongly near the threshold, hence posing a danger of it being missed out if the error bar encompasses the minute difference between the two configurations. These ambiguous pair poles are not accommodated when we use the effective range expansion.

Moving forward, we plan to use the present formalism to improve the deep learning extraction of pole configurations started in \cite{sombilloPhysRevD.102.016024, Sombillo2021classfyng, SombilloPhysRevD.104.036001}. The independence of the poles used in generating the line shape ensures that no specific trajectory is preferred to reach a particular pole configuration. Together with the vast parameters that a deep neural network can provide and its ability to generalize beyond the training dataset, one can then extract a non-biased interpretation of near-threshold enhancements.

\section*{Acknowledgment}
This work was funded by the UP System Enhanced Creative Work and Research Grant (ECWRG-2021-2-12R).
%This study was supported in part by MEXT as “Program for Promoting Researches on the Supercomputer Fugaku” (Simulation for basic science: from fundamental laws of particles to creation of nuclei).
%DLBS is supported in part by the DOST-SEI ASTHRDP postdoctoral research fellowship.
%YI is partly supported by JSPS KAKENHI Nos. JP17K14287 (B) and 21K03555 (C).
%AH is supported in part by JSPS KAKENHI No. JP17K05441 (C) and Grants-in-Aid for Scientific Research on Innovative Areas, No. 18H05407 and 19H05104.

% \section*{Appendix: Trajectories of poles in separable potential model}
% \label{sec:append}
% \hspace{\parindent}

%\begin{figure}[H]
%    \centering
%	\includegraphics[width=0.8\columnwidth]{BB_pole.png}
%	\caption{Elements of the $T$-matrix with an isolated pole on the $[bb]$ sheet. The gray dot indicates that the pole is on $[bb]$ sheet.}
%	\label{fig:bb}
%\end{figure}

%\begin{figure}[H]
%    \centering
%	\includegraphics[width=0.8\columnwidth]{BB_pole_amb.png}
%	\caption{Elements of the $T$-matrix with two poles on $[bb]$ sheet and one pole in $[tb]$. The gray dot indicates that the pole is on $[bb]$ sheet.}
%	\label{fig:bb_amb}
%\end{figure}

%\begin{figure}[H]
%	\includegraphics[width=0.8\columnwidth]{TB_pole.png}
%	\caption{Elements of the $T$-matrix with an isolated pole on the $[tb]$ sheet.}
%	\label{fig:tb}
%\end{figure}

%\begin{figure}[H]
%	\includegraphics[width=0.8\columnwidth]{TB_pole_amb.png}
%	\caption{Elements of the $T$-matrix with two poles on the $[tb]$ sheet and a single pole on the $[bb]$ sheet.}
%	\label{fig:tb_amb}
%\end{figure}

\bibliography{mybib}

\end{document}